\documentclass[entropy,article,accept,moreauthors, 10pt,a4paper]{mdpi}

\usepackage{tensor}
\usepackage{pdfpages}
\usepackage{subcaption}
\DeclareMathOperator{\arctanh}{arctanh}

\firstpage{1}
\articlenumber{x}
\doinum{10.3390/------}
\pubvolume{xx}
\pubyear{2016}
\copyrightyear{2016}
\externaleditor{Academic Editor: name}
\history{Received: date; Accepted: date; Published: date}
 \theoremstyle{mdpi}
 \newcounter{thm}
 \newcounter{ex}
 \newcounter{re}
 \theoremstyle{mdpidefinition}

\Title{GENERAL BULK-VISCOUS SOLUTIONS AND ESTIMATES OF BULK VISCOSITY IN THE COSMIC FLUID}

\Author{Ben David Normann $^{1}$ and Iver Brevik$^{2}$*}
\AuthorNames{Firstname Lastname, Firstname Lastname and Firstname Lastname}

\address{%
$^{1}$ \quad Department of Physics, Norwegian University
of Science and Technology, N-7491 Trondheim, Norway\\
$^{2}$ \quad Department of Energy and Process Engineering, Norwegian University of Science and Technology, N-7491 Trondheim, Norway}

\corres{Correspondence: iver.h.brevik@ntnu.no}

\abstract{We derive a general formalism for bulk viscous solutions of the energy-conservation-equation for $\rho(a,\zeta)$, both for a single-component and a multicomponent fluid in the Friedmann universe. For our purposes these general solutions become valuable in estimating order of magnitude of the phenomenological viscosity in the cosmic fluid at present. $H(z)$ observations are found to put an upper limit on the magnitude of the modulus of the present day bulk viscosity. It is found to be $\zeta_0\sim 10^6~$Pa s , in agreement with previous works.  We point out that this magnitude is acceptable from a hydrodynamic point of view. Finally, we bring new insight by using our estimates of $\zeta$ to analyse the fate of the future universe. Of special interest is the case $\zeta \propto \sqrt{\rho}$ for which the fluid, originally situated in the quintessence region, may slide through the phantom barrier and inevitably be driven into a big rip. Typical rip times are found to be a few hundred Gy. }


\PACS{J0101}
\begin{document}

%%%%%%%%%%%%%%%%%%%%%%%%%%%%%%%%%%%%%%%%%%
%% Sections that are not mandatory are listed as such. The section titles given are for Articles. Review papers and other article types have a more flexible structure.

%% Only for the journal Gels: Please place the Experimental Section after the Conclusions

%%%%%%%%%%%%%%%%%%%%%%%%%%%%%%%%%%%%%%%%%%
%\setcounter{section}{-1} %% Remove this when starting to work on the template.

\section{Introduction}
\label{ChaptIntro}
\noindent Recent years have witnessed  a considerable interest in theories of the dark energy cosmic fluid in the late universe. With present time defined as  $t=0$ this  means the region $t>0$.  The interest in this topic is very natural, in view of current observations of the equation-of-state parameter, commonly called $w$. From the 2015 Planck data,  Table 5 in Ref.~\cite{ade15}, we have $w=-1.019^{+0.075}_{-0.080}$. Writing the equation of state in the usual homogeneous form
\begin{equation}
p=w\rho, \quad w={\rm constant} \equiv -1+\alpha, \label{1}
\end{equation}
with the parameter $w$ here assumed constant for simplicity,
we see that the value of $\alpha$ lies between two limits,
\begin{equation}
\alpha_{\rm min}=-0.099, \quad \alpha_{\rm max}=+0.056. \label{2}
\end{equation}
It is thus quite conceivable that the cosmic fluid can be regarded as a dark energy fluid (the region $-1<w<-1/3$ is called the quintessence region, while $w<-1$ is the phantom region). Observing that the dark energy fraction  is so dominant, about $70\%$, it has, for the future universe, been common to describe the cosmic fluid as a one-component fluid \cite{stefancic05, Gorbunova, brevik06, nojiri05, Frontiers, brevik15b, disconzi15, nojiri04} for instance in the search for future singularities. For some years  it has been known that if the cosmic fluid starts out from some value of $w$ lying in the phantom region, it will encounter some form of singularity in the remote future. The most dramatic event is called the big rip, in which the fluid enters into a singularity after a finite time $t_s$ given by \cite{caldwell03,nojiri03,nojiri04}
\begin{equation}
t_{\rm s}=\frac{2}{|\alpha|\theta_0}=   \frac{2}{|\alpha|\sqrt{24\pi G\rho_0}}, \quad \theta_0=3H_0, \label{3}
\end{equation}
$\theta_0$ being the scalar expansion and $H_0$  the Hubble parameter at present time. There exist also softer variants of the future singularity where the singularity is not reached until an infinite time, called the little rip \cite{frampton11,brevik11,brevik12},  the pseudo rip \cite{frampton12}, and the quasi rip \cite{wei12}.

In various previous contexts the effects of relaxing the constancy of $w$ have been investigated, assuming instead that this quantity depends on $\rho$,
\begin{equation}
p=w(\rho)\rho, \label{4}
\end{equation}
with
\begin{equation}
w(\rho)=-1+\alpha \,\tilde{\rho}^{\gamma-1}, \quad {\rm with} \quad \tilde{\rho}=\rho/\rho_0.         \label{5}
\end{equation}
where $\alpha$ and $\gamma$ are nondimensional constants (subscript zero refers to $t=0$). The ansatz (\ref{5}) is meant to apply regardless of whether the fluid is in the quintessence or the phantom region. On physical grounds we expect that $\gamma \ge 1$. If $\gamma=1$, Eq.~(\ref{5}) reduces to $w=-1+\alpha$, i.e. the same as Eq.~(\ref{1}). If $\gamma >1$, and the fluid develops as a phantom fluid, then the influence from the density on the pressure becomes strongly enhanced near the big rip where $\rho \rightarrow \infty$. The form (\ref{5})  has previously been investigated in Refs.~\cite{nojiri05,Gorbunova,stefancic05,Frontiers}.

In the present work we will not consider the generalization contained in Eq. (\ref{5}) further.  Instead we will generalize, at least in principle, by allowing for a multi-component fluid.  There are several earlier works in this direction; cf., for instance, Refs.~
\citep{wang14,velten13,bamba12,elizalde14,brevik15,brevikTimoshkin16}.  Such a model means that the total energy density is written as  $\rho=\sum_i\rho_i$. Treating ordinary matter and dark matter on the same footing, we have, according to the $\Lambda$CDM model,  $\,\Omega_{0\rm m}+\Omega_\Lambda+\Omega_{0\rm K}=1$ (actually, $\Omega_{0\rm K}$ is  a one parameter extension of the base model).  Here $\Omega_{0\rm i}$ denotes the relative density of component $i$ at present. $i=\rm m$ denotes matter (mainly dark matter), and $i=\rm K$ includes the curvature contribution. $\Lambda$ is the cosmological constant.  Again referring to the Planck data, Table 4 in \citep{ade15}, we have  $\Omega_{\Lambda}=0.6911\pm 0.0062$, $\Omega_{0\rm m}=0.3089\pm 0.0062$, when $68\%$ intervals are considered. This already adds up to 1, and the remaining one parameter extension $|\Omega_{0\rm K}|< 0.005$ will for the present purposes be neglected. We shall, however, briefly consider the one-parameter extension of radiation, $\Omega_{0\rm r}$.

 As a second generalization we will take into account the bulk viscosity of the cosmic fluid. As is known, there exists also a second viscosity coefficient, the shear viscosity~\cite{floerchinger15}, to be considered in the general case when one works to the first order deviations from thermal equilibrium. The shear coefficient is  of particular importance when dealing with
flow near solid surfaces, but it can  be crucial also under
boundary-free conditions such as in isotropic turbulence (for cosmological applications, cf. Refs.~\cite{brevik12,brevik11a}).  When the fluid is spatially isotropic, the shear viscosity is usually left out, and we will make the same assumption here. Then only the
bulk viscosity  $\zeta$ remains in the fluid's energy-momentum tensor. It is notable that in recent years it has become quite common to include  viscous aspects of the cosmic fluid (readers interested in  general accounts of viscous cosmology under various circumstances may consult, for instance, Refs.~\cite{weinberg71,weinberg72,zimdahl96,BrevikGron13,bamba15}).

We will make the following ansatz for the bulk viscosity:
\begin{equation}
\zeta(\rho)=\zeta_0 \left( \frac{\theta}{\theta_0}\right)^{2\lambda}= \zeta_0 \,\left(\frac{\rho}{\rho_0}\right)^{\lambda}, \label{6}
\end{equation}
where $\lambda \geq 0$, and $\zeta_0$ is the present viscosity. The above ansatz, for some power of $\lambda$, has often been used in the literature, both for the early universe \citep{murphy73, barrow86, li10, campo07}  and for the later universe ~\citep{Gorbunova, brevik06, nojiri05a, paolis10, Frontiers, cardenas15}. The two most actual values for $\lambda$ are $\lambda=\frac{1}{2}$ whereby $\zeta \propto \theta \propto \sqrt \rho$, and $\lambda=1$ whereby $\zeta \propto \theta^2\propto \rho$. Again considering the case of a dark fluid, we see that Eq.~(\ref{6}) predicts $\zeta \rightarrow \infty$ near the big rip where $\theta \rightarrow \infty$. In some of the previous literature mentioned above, both Eq. (\ref{6}) and Eq. (\ref{5}) are assumed at the same time, with $\gamma-1 =\lambda$. As mentioned,  Eq.~(\ref{5}) is however not assumed in the present work; for clarity we take $w=-1+\alpha$ throughout.

When dealing with the future universe, one needs to have information about the value of the present-day viscosity $\zeta_0$, and the coefficient $\lambda$. To achieve this, one has to take into account observations about the past universe (in our notation $t<0$). We will work out below general solutions from which estimates can be given  for  these two quantities.  Especially the  magnitude of $\zeta_0$ is of interest, as little seems to be known about this quantity from before. We intend to come back to an analysis  of these  general solutions in a later paper.

 It is to be borne in mind that the inclusion of a bulk viscosity is done only on a phenomenological basis. There might be fundamental reasons for the viscosity, based upon kinetic theory, but this is a different topic, and readers interested in such a line of research should consider for instance \cite{zimdahl96, horn16}. From an analogy with standard hydrodynamics a phenomenological approach  is obviously  natural.

Let us now  write down the standard FRW metric, assuming zero spatial curvature, $k=0$,
\begin{equation}
ds^2=-dt^2+a^2(t)\,d{\bf x}^2. \label{7}
\end{equation}
 The energy-momentum tensor for the whole fluid is
\begin{equation}
\label{MCF_EnMomTens}
T_{\mu\nu}=\rho U_\mu U_\nu+(p-\theta\zeta)h_{\mu\nu},
\end{equation}
where $h_{\mu\nu}=g_{\mu\nu}+
U_\mu U_\nu$ is the projection tensor. In co-moving coordinates ($U^0=1$, $U^{i}=0$), and with the metric \eqref{7}, Einstein's equation gives the two Friedmann equations
\begin{equation}
\theta^2=24\pi G\rho, \label{MCF_Fr1}
\end{equation}
\begin{equation}
\dot{\theta}+\frac{1}{2}\theta^2=-12\pi G\left(p-\zeta(\rho)\theta\right), \label{MCF_Fr2}
\end{equation}
where $\rho$ denotes the cosmological fluid as a whole.  By Eq.~\eqref{MCF_EnMomTens} the  conservation equation for energy and momentum  becomes for the overall fluid
\begin{equation}
\tensor{T}{^{\mu\nu}_{;\nu}}=0\phantom{000}\Rightarrow\phantom{000}\dot{\rho}+(\rho+p)\theta=\zeta\,\theta^2 ~~{\rm
when}~~
 \mu=0.
\label{MCF_EEq}
\end{equation}

The following point ought here to be noted. If we simply impose the conservation equation $ {T_i^{\mu\nu}}_{;\nu}=0$ for the matter subsystem {\it i=m}, we will get, for $\mu=0$,
\begin{equation}
\dot{\rho}_m+(\rho_m+p_m)\theta=\zeta_m\theta^2, \label{ekstra1}
\end{equation}
with $\zeta_m$ referring to the matter. Compare this with the balance equations for energy following from the assumption about an interacting system consisting of matter and dark energy (de) fluid,
\begin{equation}
\dot{\rho}_{\rm m}+(\rho_{\rm m}+p_{\rm m})\theta=Q, \label{ekstra2}
\end{equation}
\begin{equation}
\dot{\rho}_{\rm de}+(\rho_{\rm de}+p_{\rm de})\theta=-Q, \label{ekstra3}
\end{equation}
where $Q$ is the energy source term. This is actually the way in which the coupling theory is usually presented (cf., for instance, the recent references \cite{brevikTimoshkin16,fay16}). Comparison between Eqs.~(\ref{ekstra1}) and (\ref{ekstra2}) shows that the coupling is in our case essentially the viscosity. This suggests that the viscosity should preferably  be taken to depend on the fluid as a whole, thus $\zeta=\zeta(H), \zeta=\zeta(\rho),$ or $\zeta=\zeta(z)$ ($z$ being the redshift), instead of being taken as a function of the fluid components.

 Our idea  will now be to   develop a general formulation  for the  viscous fluid, and  to compare the theoretical predictions with measurements. As mentioned, similar  approaches have been applied in Refs.~\citep{wang14} and \citep{velten13}, but not in the  general way here considered. Our main reason for developing this framework is, as mentioned, to  study  the future universe. The formalism as such is applicable to the past as well to the future universe,  and we need to use observations from the past universe in order to get an idea about its future development. We intend also to relate   various models presented  in the literature to each other.
\bigskip

\textit{Section  2} contains a central part of our work, as general bulk-viscous solutions are presented for $\zeta(z)$  and $\zeta(\rho)$, respectively. We justify our approach and present the underlying assumptions. In \textit{section 3} we implement a definite model with the theoretical framework worked out in the foregoing sections. The section also contains some simple non-linear regressions for three different models of the bulk viscosity. In \textit{section 4} we discuss our results, with emphasis on the model where $\zeta \propto \sqrt \rho$.  Magnitudes of the viscosity suggested so far in the literature, are considered. Finally, on the basis of the obtained value for the viscosity, we return in \textit{section 5} to
 the future universe. In particular, we estimate the time needed to run into the big rip singularity.

\section{General solutions, assuming $\zeta=\zeta(\rho)$}
\label{Chap_GenSol}
\noindent In the present section we let the viscosity be dependent on the overall density $\rho$ of the cosmic fluid, i.e. $\zeta(\rho)$. We  start by solving Eq.~\eqref{MCF_EEq} (restated below) with respect to $\rho(a,\zeta)$. Thereafter this solution is used in the first Friedmann equation \eqref{MCF_Fr1} to find $E(a,\zeta)$, where $E=H/H_0$ is the dimensionless Hubble parameter. We introduce the definition
\begin{equation}
\label{C_B}
B\equiv12\pi G\zeta_0,
\end{equation}
as a useful abbreviation, where $\zeta_0$ is the present viscosity  (divide by $1/c^2$ to convert to physical units). This definition   differs  from that found in \citep{wang14} only by the omission of $T_0^{\delta}$, since we do not consider temperatures in this approach. In physical units the dimension of $B$ is the same as that of the Hubble parameter, $[B]=[H_0]=$s$^{-1}$. One may for convenience express $B$ in the conventional astronomical units, km s$^{-1}$Mpc$^{-1}$.  If we denote this quantity as $B$[astro.units], we  obtain

\begin{equation}
\label{C_B_to_Zeta}
\zeta_0 = B{\rm [astro.units]}\times 1.15\times 10^6 \text{ Pa s},
\end{equation}
which is a useful conversion formula. Now consider the energy conservation equation \eqref{MCF_EEq} following from \eqref{MCF_EnMomTens},
\begin{equation}
\label{C_EnergyEq}
a\partial_a\rho(a)+3[\rho(a)+p]=3\zeta(\rho)\theta
\end{equation}
when rewritten in terms of the scale factor $a$. Evidently  the viscosity here refers to the {\it fluid as a whole}. By the inclusion of $\zeta$ in the equation~\eqref{MCF_EnMomTens} for $\tensor{T}{^{\mu\nu}}$ we have ensured a divergence-free total energy-momentum tensor  ${T^{\mu\nu}}
_{;\nu}=0$ by construction. But the interpretations of the phenomenologically included $\zeta$ is to extent open, as we have already anticipated. It depends essentially on whether we take the fluid to be a one-component, or a multicomponent, system  (cf. a closer discussion in Appendix \ref{App:A}). This is a matter of physical interpretation, and does not need to be specified for the purposes of the present section. We do not here require that  $\tensor{T}{_i^{\mu\nu}_{;\nu}}=0$ for each component $i$. See also the brief discussion on this point in Appendix \ref{App:B}.

We have so far made no assumption about the form of $\rho(a)$. For a general multicomponent fluid with an arbitrary number of components, we can however write $\rho=\sum_{i}\rho_i$,
where the sum goes over an arbitrary number of components. First, if there is no viscosity, we have
 \begin{equation}
\label{C_GeneralEoS_CosmicFluid}
p=\sum_{i}w_i\rho_i,\phantom{0000}\textbf{assumption 1},
\end{equation}
 which means that each component $i$ contributes linearly to the overall pressure $p$. In this case the energy-conservation-equation is easily verified to have the homogeneous solution (i.e. $\zeta=0$)
\begin{equation}
\label{C_HomSol_EnEq}
\rho_{\rm h} (a)=   \sum_i \rho_{ {\rm h} i} (a)=       \sum_{i}\rho_{0i}a^{-3(w_i+1)},
\end{equation}
where $\rho_{0i}$ are the present densities $(a_0=1)$. Thus in the absence of viscosity  the overall fluid would evolve as \eqref{C_HomSol_EnEq}. Now including viscosity, we   let the general solution be a sum of a homogeneous and a particular one, so that
\begin{equation}
\label{C_GenSol_EnEq_1}
\rho(a)=\sum_{i}\rho_{{\rm h}i}(a)+\rho_{\rm p}(a, \zeta)=\sum_{i}\rho_{{\rm h}i}(a)\left[1+u_i(a, \zeta)\right]=\sum_{i}\rho_{0i}a^{-3(w_i+1)}\left[1+u_i(a, \zeta)\right],
\end{equation}
where $u_i(a)$ are functions to be determined by substituting equation \eqref{C_GenSol_EnEq_1} for $\rho$ in the energy-conservation-equation \eqref{C_EnergyEq}. Doing so, we find the differential equation

\begin{equation}
\label{C_DiffEq_u1}
\sum_i\rho_{\text{h}i}(a)\frac{\partial u_i(a, \zeta)}{\partial a}=3\frac{\zeta(\rho)}{a}\theta.
\end{equation}
Inserting  $\theta$ from the first Friedmann equation \eqref{MCF_Fr1} we find
\begin{equation}
\label{C_DiffEq_u2}
\sum_i\rho_{\text{h}i}(a)\frac{\partial u_i(a, \zeta)}{\partial a}=9\frac{\zeta(\rho)}{a}\sqrt{\frac{8\pi G}{3}\sum_i\rho_{\text{h}i}(a)
\left[1+u_i(a, \zeta)\right]}.
\end{equation}
This equation, as it stands,  is not particularly useful. In principle, one might solve it for one component $u_i(a, \zeta)$, but since the equation is non-linear in $\rho$, the superposition principle cannot be used to find the solution for a multicomponent fluid with density $\rho(a)$. This would mean that different $u_i$s must be calculated, since the viscosity effect would be different for the different components. We will follow a simpler approach; by noting that the above equation can be solved if all the $u_i(a, \zeta)$s are equal; $u_i(a, \zeta)\rightarrow u(a, \zeta)$. In this way, the non-linearity of \eqref{C_DiffEq_u2} in $\rho$ is avoided. Physically, this means introducing  a phenomenological  viscosity for the overall fluid. Equation \eqref{C_GenSol_EnEq_1} now becomes

\begin{equation}
\label{C_GenSol_EnEq_2}
\rho(a)= \rho_{\rm h}(a)\left[1+u(a)\right],\phantom{0000}\textbf{assumption 2}.
\end{equation}
This assumption simplifies the formalism. Note that the relative contributions of the fluid components for any redshift remain unaltered compared to the inviscid case. By the above assumption Eq.~\eqref{C_DiffEq_u2} becomes
\begin{equation}
\label{C_DiffEq_u3}
\frac{\partial u(a)}{\partial a}=9\frac{\zeta(a)}{a\rho_h(a)}\sqrt{\frac{8\pi G}{3}\rho_h(a)\left[1+u(a, \zeta)\right]}.
\end{equation}
We may on this point refer to  \citep{velten13}, where rather general remarks are made in the case of $\zeta\rightarrow\zeta(\rho_i)$. Eq.~\eqref{C_DiffEq_u3} may now be solved for $u(a,\zeta)$, if $\zeta(\rho)$ is known. Inserting our ansatz Eq.~\eqref{6} for $\zeta$ we find
\begin{equation}
\label{C_DiffEq_ZetaRho_u2}
\frac{1}{(1+u)^{\lambda+1/2}}\frac{du}{da}=\frac{9\zeta_0}{a\rho_0^\lambda}\sqrt{\frac{8\pi G}{3}}\left(\rho_{\text{h}}(a)
\right)^{\lambda-1/2},
\end{equation}
where the arguments of $u$ were suppressed for brevity. The solution is
\begin{equation}
\label{C_Sol_u_final}
\boxed{
u(z,B,\lambda)=
\begin{cases}
\displaystyle\left[ 1-\left(1-2\lambda\right)\frac{B}{H_0}\int_0^z{\frac{1}{(1+z)\sqrt{\Omega}^{1-2\lambda}}}dz\right]^{\frac{2}{1-2\lambda}}-1
& \phantom{00}\text{for}\lambda\neq \frac{1}{2},\\
\displaystyle (1+z)^{-\frac{2B}{H_0}}
& \phantom{00}\text{for}\lambda=\frac{1}{2},\\
\end{cases}
}
\end{equation}
where we have rewritten in terms of the redshift through $a=1/(1+z)$, and where the initial condition was chosen such that $\rho(z=0,\zeta=0)=\sum_i\rho_{0i}$. Also, for brevity,
\begin{equation}
\label{C_defOmega}
\Omega\equiv\sum_i\Omega_{0i}(1+z)^{3(1+w_i)}\phantom{000}\text{where}\phantom{000}\Omega_{0i}=\frac{\rho_{0i}}{\rho_c}\phantom{000}\text{and}\phantom{000}\rho_c=\frac{3\,H_0^2}{8\pi\,G}
\end{equation}
and as defined previously; $B=12\pi G\zeta_0$. By Eqs.~\eqref{C_GenSol_EnEq_2} Friedmann's first equation \eqref{MCF_Fr1} now gives the dimensionless Hubble parameter $E(z)$ as
\begin{equation}
\label{C_Ez_zetaRho}
\boxed{
E^2(z)=
\displaystyle\Omega\left[1+u(\lambda,\zeta_0)\right]}\phantom{0},
\end{equation}
where $u(\lambda, \zeta_0)$ is given by the solutions~\eqref{C_Sol_u_final}. Initial condition $E(z=0)=1$ is fulfilled. In the case of zero viscosity, Eq.~ \eqref{C_Ez_zetaRho} reduces to the first Friedmann equation (with $k=0$) on standard dimensionless form. This is as expected, and  shows that the particular solution is needed in order to account for the viscosity properly. Since the above equations are valid for any number of components, it should be possible to apply them in many different scenarios, also inflationary scenarios, for instance as a natural extension of the case studied in \cite{brevikTimoshkin16}. The general solution of the integral in Eq.~\eqref{C_Sol_u_final} is quite involved, but we solve it for the specific cases $\lambda=1/2$ and $\lambda=1$ which, as mentioned in the introduction, are among  the most popular choices. We end this section by noting that also one-fluid models, such as the kind found, for instance,  in Ref.~\cite{Frontiers}, naturally becomes a special case of our general solutions. Eq.~\eqref{C_Ez_zetaRho} presents the cases that we shall study further in the present work. But before that we shall briefly comment on theoretical aspects of the case $\zeta(z)$.

\subsection{Comments on the case $\zeta(z)$}
\noindent The energy-conservation-equation is solvable also in the case of $\zeta(z)$. A redshift dependent viscosity might be more natural in some cases, like the treatment given in \cite{wang14}. Following the same procedure as in the case $\zeta (\rho)$ presented above, one this time finds
\begin{align}
\label{C_FriedmannVisc1Ez_1}
E^2(z)=\Omega\left[1+u(\zeta)\right]
&&\text{where}&&
u(\zeta)=\left[1-\frac{B}{H_0}\int_0^z{\frac{\zeta(z)}{(1+z)\sqrt{\Omega}}}{\,dz}\right]^2-1,
\end{align}
when initial condition $E(z=0)=1$ is fulfilled. We shall not use these solutions any further in this paper.

\section{Implementing the theory with realistic universe models and determining $\zeta_0$}
\subsection{Restricting the number of components in the fluid model}
\noindent Now that the general bulk-viscous framework is in place, one may attempt at implementing specific universe models. In particular, what needs to be determined, is which components one should include in the cosmic fluid, and what kind of viscosity. In \cite{chen15} one finds Hubble parameter measurements back to redshifts $\sim 2.3$. As Table~ \ref{tab:CosmicEvolution} shows, this stretches deep into the matter dominated epoch. As is known,  at redshift $z=0.25$ dark energy becomes the main constituent. Taken all together, it is natural as a first approach to assume the universe consisting of dust ($w=0$) and a constant dark energy term ($w=-1$). With $\rho (z)\rightarrow\rho_m(z)+\rho_{\rm de}$, we find
\begin{equation}
\label{C_FriedmannVisc1Ez_ZetaConst_1}
E^2(z)=[\Omega_{\rm de}+\Omega_{0 \rm m}(1+z)^{3}]\left(1+u\right),
\end{equation}
where $u$ now is given by Eq.~\eqref{C_Sol_u_final}, wince we intend to give the viscosities as function of $\rho$. We intend in the following to give an estimate of the viscosity useful for future properties of the cosmic fluid, such as singularities like the big rip. In the previous investigations to this end, a phenomenological one-component approach has been used (cf. the introduction). Since we want to follow the well established $\Lambda$CDM model as closely as possible, we will not consider a one component fluid. However, we will assign a bulk viscosity only to the fluid as a whole. This gives a natural transition into a one-component phenomenological description of the future cosmic fluid. In the following, we shall implement the three most used cases $\zeta=\rm const$, $\zeta\propto\sqrt{\rho}$ and $\zeta\propto\rho$ in order to estimate the magnitude of the viscosity $\zeta_0$. The whole point with estimating $\zeta_0$ is in the present context to determine its impact on properties of the future cosmic fluid.
\begin{table}[H]
\centering
\begin{tabular}{llll}
\toprule
\multicolumn{4}{l}{\textbf{Cosmological Evolution}}\\
\hline
\textit{Cosmic time} & \textit{scale factor a}& \textit{Era} &  \textit{Redshifts}\\
\hline
$t=13.8$ Gy& 1 & Present & 0\\
$9.8$ Gy$<t<13.8$ Gy& $a(t)=e^{H_0 t}$ & DE dominance & -\\
$t=9.8$ Gy& 0.75 & onset of DE dominance & 0.25\\
$47$ ky$<t<9.8$ Gy& $a(t)\propto t^{2/3}$ & matter dominance & -\\
$t=47$ ky & $1.2\cdot 10^{-4}$ & onset of matter dominance & 3400\\
$t<47$ ky& $a(t)\propto t^{1/2}$ &  radiation dominance & -\\
$t=10^{-10}$ s& $1.7\cdot 10^{-15}$ & electroweak phase transition & -\\
$10^{-44}$s$<t<10^{-10}$s&$ a(t)\propto t^{1/2}$ & Possible inflation or bounce & -\\
$t<10^{-44}$& $1.7\cdot 10^{-32}$ & Planck time & $5.9\cdot 10^{31}$\\
\hline
\end{tabular}
\caption{Overview over cosmological time as function of redshift. The first three columns are based on Ref.~\citep{frampton15}; the last column contains  useful approximate redshifts.}
\label{tab:CosmicEvolution}
\end{table}

\subsection{Explicit formulae for $E(z,\zeta_0)$ obtained for the three cases $\,\zeta=$const., $\zeta\propto\sqrt{\rho}$ and $\zeta\propto\rho$}
\noindent Solving the integral in Eq. \eqref{C_FriedmannVisc1Ez_1} for $u(z)$ in the three different cases, equation \eqref{C_FriedmannVisc1Ez_ZetaConst_1} becomes
\begin{equation}
\label{FriedmannViscRho}
E(z)=
\begin{cases}
\displaystyle \sqrt{\Omega(z)}\left[1-\frac{2B}{3H_0\sqrt{\Omega_{\rm de}}}\arctanh \left(\sqrt{\frac{\Omega(z)}{\Omega_{\rm de}}}\right)+I_0\right]
& \phantom{000}\textbf{when}\phantom{000}\zeta=\rm const.,\\
\displaystyle \sqrt{\Omega(z)}(1+z)^{-\frac{B}{H_0}}
&\phantom{000}\textbf{when}\phantom{000}\zeta=\zeta_0\left(\frac{\rho}{\rho_0}\right)^{1/2},\\
\displaystyle \frac{\sqrt{\Omega}}{\sqrt{1+\frac{2B}{3H_0}\left[\sqrt{\Omega}\left(1-\frac{\sqrt{\Omega_{\rm de}}}{\sqrt{\Omega}}\arctanh\sqrt{1+\frac{\Omega_{\rm 0m}}{\Omega_{\rm de}}(1+z)^3}\right)\right]+C}}
&\phantom{000}\textbf{when}\phantom{000}\zeta\propto\rho.\\
\end{cases}
\end{equation}
where we have rewritten the expressions in terms of relative densities. The definition \eqref{C_B} has also been used. The integration constants are readily determined by the initial condition $E(z=0)=\Omega(z=0)\equiv\Omega_0=1$.

\subsection{Data fitting}
\noindent Before we go on to discuss the future universe, we need an estimate of the magnitude of the viscosity. For the present purposes, an estimate of order of magnitude suffices, and hence we will apply a simple procedure. From the formulae in the previous section, we are able to estimate an upper limit on the magnitude of the modulus of $\zeta_0$. This was done by minimizing
\begin{equation}
\label{statistics}
\chi_{\rm H}^2(H_0,\zeta)=\sum_{i=1}^{N}\frac{\left[H^{\rm th}(z_i; H_0,\zeta)-H^{\rm obs}(z_i)\right]^2}{\sigma_{{\rm H},i}^2},
\end{equation}
where $N$ is the number of data points, $H^{\rm th}(z_i)$ is the theoretical Hubble parameter value at redshift $z_i$, $H^{\rm obs}(z_i)$ is the observed value at redshift $z_i$ and $\sigma_{{\rm H},i}^2$ is the variance in observation $i$. To the best of knowledge, \citep{chen15} contains the most up-to-date set of independent $H(z)$ observations. To estimate orders of magnitude from the prescriptions found therein, we minimize Eq.~\eqref{statistics} through a  non-linear least square procedure. Table~\ref{tab:ModelSummary} compares the fit of the different assumptions made for $\zeta$. In the most recent Planck data ~\citep{ade15}, Table 4, one finds the values $H_0=67.74\,\text{ km}\,\text{s}^{-1}\,\text{Mpc}^{-1}\phantom{00}\text{,}\phantom{00}\Omega_{0\text{m}}=0.3089,\phantom{00}\Omega_{\rm de}=0.6911.$ These values were used in our regression.
\begin{table}[H]
\centering
\begin{tabular}{lccr}
\toprule
\multicolumn{4}{c}{\textbf{Summary of Model fitting}} \\
\midrule
Model for $\zeta$	&	Adjusted R$^2$	&	Fit-value for $B$  	&	$95\%$CI\\
&$[-]$&$\rm(km\phantom{0}s^{-1}\phantom{0}Mpc^{-1}$)& \\
\hline
$\zeta=$const.&0.9601& 0.6873&(-2.788, 4.163)\\
  $\zeta\propto\rho^{1/2}\propto H$&0.9604&0.7547&(-1.706, 3.215)\\
$\zeta\propto\rho\propto H^2$&0.9609&0.5906&(-0.8498, 2.031)\\
\hline
\end{tabular}
\caption{Results of the different models that have been compared with observations.}
\label{tab:ModelSummary}
\end{table}
A lot more could be done to obtain accurate estimates of $\zeta_0$. Especially by including different data sets. However, this is not necessary for our purposes, and we leave it for future investigations.

\begin{figure}[H]
\centering
   \includegraphics[trim=5cm 9.7cm 5cm 10cm, width=0.7\linewidth]{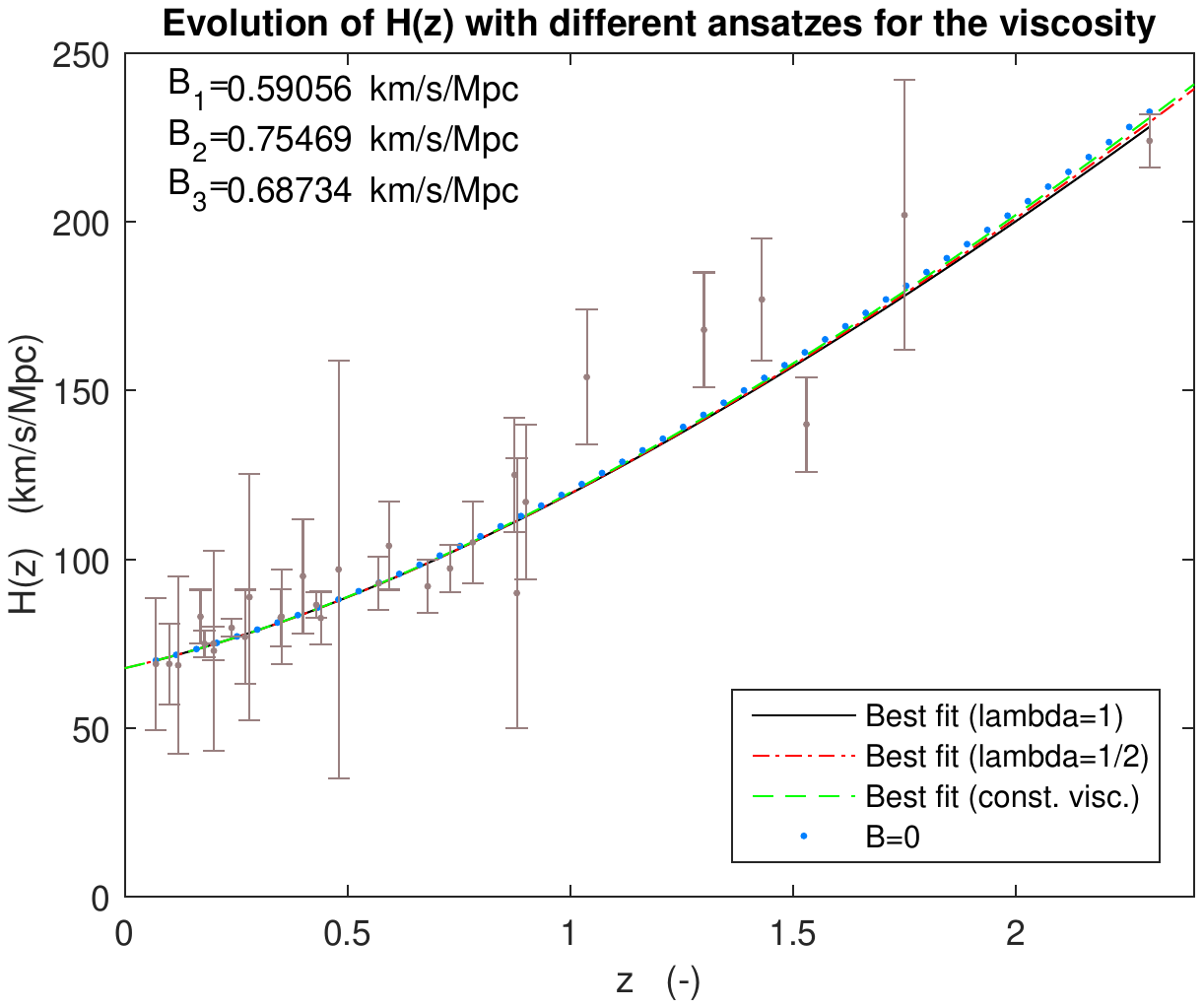}
\label{fig:BF_All}
\caption{Friedmann's first equation for $H(z)$ [km s$^{-1}$Mpc$^{-1}$] with three different ansatzes for the viscosity. The fit-values used for $B$ are (i) ~$B_1=\,0.590$ km s$^{-1}$Mpc$^{-1}$ for $\zeta\propto\rho$ ~(solid, black line), ~(ii) ~$B_2=\,0.755$ km s$^{-1}$Mpc$^{-1}$ for $\zeta\propto\sqrt\rho$ ~(stipled, dotted red line) and (iii) ~$B_3=0.687$ km s$^{-1}$Mpc$^{-1}$ for constant viscosity ~(stipled, green line). Dotted blue line gives the corresponding evolution for $B=0$ (no viscosity).}
\end{figure}

%----------------------------------------------------------------

\section{Discussion and further connection to previous works}
\label{Discussion}
\noindent Additional information should be taken into account in order to help deciding  between the three cases $\zeta_0$, $\,\zeta\propto\sqrt{\rho}\,$ and $\,\zeta\propto\rho$ discussed above. In this section we make comments on this point and also on the expected magnitude of $\zeta_0$.

\subsection{The evolution of $\zeta$}
\noindent The three functional forms implemented in this paper appear to be widely accepted  (cf. the Introduction). As Ref.~\citep{cardenas15} formulates  it, the most common dependencies $\zeta \propto \rho$ and $\propto \sqrt{\rho}$ are chosen because they lead to well known analytic solutions. Any attempt at extrapolating the theory into the future will involve knowledge about  the functional form of $\zeta$. In the following we consider the option $\zeta \propto\sqrt{\rho}$ and wish to point out that this form for $\zeta$ -- at least from a dynamical point of view --  has the characteristic property that it is subject to  multiple possibilities of interpretation. Going back to the energy-conservation equation, Eq.~\eqref{C_EnergyEq} and inserting $\zeta=\zeta_0\theta /\theta_0$ we can move the right-hand-side over to the left hand side and find
\begin{equation}
\label{C_EnergyEq_rest2}
a\partial_a\rho(a)+3[\rho(a)+p]=\frac{3\zeta_0}{\theta_0}\theta^2\phantom{000}\rightarrow\phantom{000}\sum_i \left[a\partial_a\rho_i(a)+3(1+w_i-\frac{2B}{\theta_0})\rho_i \right]=0.
\end{equation}
$p=\sum_iw_i\rho_i$ and $B=12\pi G\zeta_0$ are used as before. From the above equation it is clear that we dynamically could obtain the same result by shifting each equation-of-state parameter $w_i$ such that
\begin{equation}
\label{omegaShift}
w '=w-\frac{2B}{\theta_0},
\end{equation}
where $w_i'$  is the new equation-of-state parameter. This property is pointed out from another perspective also in e.g. Refs.~\cite{nojiri05} and \cite{velten13}. The result is seen to confirm with  the solution  of Eq. \eqref{C_Ez_zetaRho}, with $\lambda=1/2$. Note that $B\sim 1$ km s$^{-1}$Mpc$^{-1}$ would correspond to a shift in $w$ of $\sim 0.01$ according to the above equation. By interpreting the viscosity as a result of interplay between various fluid components with homogeneous equations of state (treated in more detail in Ref.~\cite{zimdahl96}), we show in Appendix \ref{App:A} that one is lead to a phenomenological viscosity of the form
\begin{equation}
\label{EfectiveVisc}
\zeta=\frac{H_0}{8\pi G}\left[w-w_1\frac{\Omega_1(z)}{\Omega(z)}-w_2\frac{\Omega_2(z)}{\Omega(z)}\right]\sqrt{\Omega(z)},
\end{equation}
where $\Omega(z)$ includes all fluid components.  This expression accounts for the phenomenological viscosity resulting when two of the components ($i=1,2$) in a multi-component fluid are seen as one single fluid component with a single equation-of-state parameter $w$. Note the functional form of $\zeta$ in the above equation: if the square bracket is well approximated by a constant, the functional form of $\zeta$ approaches $\zeta\propto \sqrt{\Omega(z)}\propto\sqrt{\rho}$. Inserting $B\sim 1$ km s$^{-1}$Mpc$^{-1}$ and the corresponding $w=0.01$ we find such a regime for redshift values $-1<\,z\,<\,1$ when extending the base $\Lambda$CDM model by including radiation and baryons  as \textit{one} effective matter/radiation component ($\rho_{\rm mr}$). This means that we now deal with a phenomenological fluid consisting of three components $\rho_{\rm de}$ ,$\rho_{\rm dm}$ and $\rho_{\rm mr}$. The kind of viscosity that we here consider, originating from lumping two or more components together, may be simplifying though obviously phenomenological. However, it is of definite interest in making predictions for the future universe, which we will do in a later section.

\subsection{The magnitude of $\zeta_0$}
\noindent The magnitude is of comparable size for all the three functional forms here tested. Using Eqs.~(\ref{C_B}) and (\ref{C_B_to_Zeta}) for the $B$-values listed in Table~\ref{tab:ModelSummary}, we seem to be on safe grounds by saying that $\zeta_0 < 10^7 \text{Pa s}$. This is largely in agreement with Ref.~\cite{wang14} and also Ref.~\citep{velten12}, wherein bulk-dissipative dark matter is considered. As pointed out in the same paper,  this is $10$ orders of magnitude higher than the bulk viscosity found in for instance water at atmospheric pressure and room temperature. We mention also the even better agreement with the conservative estimate recently given in Refs.~\cite{brevik15a} and \cite{brevik15b}, where the interval $10^4~{\rm Pa~ s} <\zeta_0 < 10^6~\rm{Pa~ s}$ is found. Also Ref.~\citep{sasidharan15}, seems to find a bulk viscosity $\zeta_0\sim 10^7~\rm{Pa~ s}$ in the case of bulk viscous matter when constant viscosity $\zeta_0$ is considered. Since in fluid mechanics the viscosity coefficients appear in connection with first order modification to thermodynamical equilibrium, it becomes natural to expect that  the pressure modification caused  by the bulk viscosity should  be much smaller than the equilibrium pressure. Using the critical density ($\rho_{\rm c}\sim 10^{-26}$ kg m$^{-3}$) as a measure of the present day pressure in the universe, and estimating  the equation-of-state parameter  $w$ to be  of order  unity, the above restriction reduces to
\begin{equation}
\label{C_EstPressure_1}
|p| =|w\rho| \gg |\zeta_0\theta|\phantom{000}\rightarrow\phantom{000}|\zeta_0 |\ll 10^{8}\text{Pa s}
\end{equation}
in SI units. $p$ here means the pressure in the overall  fluid, and $\rho$ is the overall density. This result shows that the viscosity coefficient actually can be extremely high compared to the intuition given by kinetic theory applied on atomic and molecular scales. Our regression found the upper limit on the magnitude of the modulus of the present day viscosity to be $\zeta_0 \sim 10^{6}\text{ Pa s}$, consistent with thermodynamics and $1\%$ of the estimated equilibrium pressure.

%%%%%%%%%%%%%%%%%%%%%%%%%%%%%%%%%%%%%%%%%%%%%%%%%%%%%

\section{Late universe: Calculation of the rip time}
\noindent Armed with the above information we can now make a quantitative calculation of the future big rip time, based upon a chosen model for the bulk viscosity. As before we let $t=0$ refer to the present time, and we shall in this section adopt the formulation for which $\zeta=\zeta(\rho)$. As before $\rho$ here refers to the cosmological fluid as a whole. Since we are looking at the future universe, we shall in this section go back to the much studied one-component model discussed in chapter I. Using Friedmann's equations and the energy conservation under the assumptions $k=0$ and $\Lambda=0$ we obtain the following governing equation for the scalar expansion $\theta$ \cite{brevik15b,Gorbunova}:
\begin{equation}
\dot{\theta}+\frac{1}{2}\alpha \theta^2-12\pi G \zeta(\rho)\theta=0,
\end{equation}
which can be rewritten in terms of the density as
\begin{equation}
\dot{\rho}+\sqrt{24\pi G}\,\alpha \rho^{3/2}-24\pi G\zeta(\rho)\rho=0.
\end{equation}
The solution is
\begin{equation}
t=\frac{1}{\sqrt{ 24\pi G}}\int_{\rho_0}^\rho \frac{d\rho}{\rho^{3/2}\left[ -\alpha +\sqrt{24\pi G}\,\zeta(\rho)/{\sqrt \rho}\right]}. \label{governingequation}
\end{equation}

We will henceforth consider two models for the bulk viscosity:

\bigskip

\noindent {\it Case 1:} $\zeta$ equal to a constant.  We  put for definiteness the value of  $\zeta$ equal to its present value,
\begin{equation}
\zeta= \zeta_0=10^5~{\rm Pa~s},
\end{equation}
i.e., the mean of the interval given previously in section 5.2. It corresponds to the viscosity time ( in dimensional units)
\begin{equation}
t_{\rm c}=\frac{c^2}{12\pi G \zeta_0}=3.58\times 10^{20}~\rm s.
\end{equation}
The rip time becomes in this case \cite{brevik15b}
\begin{equation}
t_{\rm s}=t_{\rm c}\ln \left( 1+\frac{2}{|\alpha|\theta_0t_{\rm c}}\right),
\end{equation}
where we have taken into account that in this case $\alpha$ has to be negative to lead to a big rip. For definiteness we choose
\begin{equation}
\alpha=-0.05,
\end{equation}
which is a reasonable negative value according to experiment; cf. Eq.~(\ref{2}). With $\theta_0=6.60\times 10^{-18}~$s$^{-1}$  we then get
\begin{equation}
t_{\rm s}=6.00\times 10^{18}~{\rm s}= 190~{\rm Gy}, \label{constantzeta}
\end{equation}
thus much larger than the age $13.8~$Gy of our present universe.

\bigskip
\noindent {\it Case 2: $\zeta \propto \sqrt{\rho}$}. We take
\begin{equation}
\zeta(\rho)=\zeta_0\sqrt{\tilde{\rho}}, \quad \tilde{\rho}=\rho/\rho_0,
\end{equation}
with $\zeta_0$ the same as above. From Eq.~(\ref{governingequation}) we then get
\begin{equation}
t=\frac{1}{\sqrt{24\pi G}}\,\frac{2}{-\alpha+ \zeta_0 \sqrt{24\pi G/\rho_0}}\left(\frac{1}{\sqrt{\rho_0}}-\frac{1}{\sqrt \rho}\right),
\end{equation}
The remarkable property of this expression, as pointed out already in Ref.~\cite{Gorbunova}, is that it permits a big rip singularity even if the fluid is initially in the quintessence region $\alpha >0$. The condition is only that
\begin{equation}
-\alpha+ \zeta_0 \sqrt{24\pi G/\rho_0} >0.
\end{equation}
If this condition holds, the universe runs into a singularity ($\rho=\infty$) at a finite rip time
\begin{equation}
t_{\rm s}= \frac{1}{\sqrt{24\pi G \rho_0}}\,\frac{2}{-\alpha+ (\zeta_0/c^2) \sqrt{24\pi G/\rho_0}},
\end{equation}
here given  in dimensional units. Identifying $\rho_0$ with the critical energy density $\rho_c=2\times 10^{-26}~$kg/m$^3$ (assuming the conventional $h$ parameter equal to 0.7), we can write the rip time in the form
\begin{equation}
t_{\rm s}=\frac{2}{-\alpha +0.0056}\times 10^{17}~\rm s,
\end{equation}
which clearly shows the delicate dependence upon $\alpha$. If the universe starts from the quintessence region, it may run into the big rip if $\alpha <0.0056$, thus very small. If the universe starts from the phantom region, it will always encounter the singularity. In the special case when $\alpha=0$ we obtain $t_{\rm s}=3.6\times 10^{19}~$s, thus even greater than the previous expression (\ref{constantzeta}) for the constant viscosity case. If $\alpha=-0.05$ as chosen above, we find $t_{\rm s}=3.59\times 10^{18}~$ s=114 Gy.

\section{Conclusion}
\noindent We may summarize as follows:

\noindent $\bullet$ The main part of this paper contains a critical survey over solutions of the energy-conservation-equation for a viscous, isotropic Friedmann universe having zero spatial curvature, $k=0$. We
assumed the equation of state in the homogeneous form $p=\sum_i w_i\rho_i$, with $w_i=$ constant for all components in the fluid. With $\rho$ meaning the energy density and $\zeta$ the bulk viscosity we focused on three options: (i) $\zeta=$const, (ii) $\zeta \propto \sqrt{\rho},$ and (iii) $\zeta \propto \rho$.  We here made use of  information from various experimentally-based sources; cf. \citep{chen15},  \cite{wang14}, and   others. Our analysis was kept on a general level, so  that previous theories, such as that presented in Ref.~\cite{Frontiers} for instance, can be considered as a special case. We also mentioned the potential to include $\zeta (z)$ cases, and component-dependent cases $\zeta_i(\rho_i)$, such as those treated in for instance Refs.~\cite{wang14} and \cite{velten13}. Note that our  solutions also have the capability to include component extensions of the base $\Lambda$CDM model, such as inclusion of radiation. This was so because we assumed a general multicomponent fluid.

\noindent $\bullet$ A characteristic property as seen from the figure is that  the differences between the  predictions from  the various viscosity models  are relatively small. It may be surprising that even the simple ansatz $\zeta=$constant reproduces experimental data quite well. These models however tend to underpredict $H(z)$ for large redshifts. In the literature, the ansatz    $\zeta \propto \sqrt{\rho}$,  is  widely accepted.

\noindent $\bullet$ As for the {\it magnitude} of the bulk viscosity $\zeta_0$ in the present universe we found, on the basis of various sources, that one hardly do better than restricting $\zeta_0$ to lie within an interval. We suggested the interval to extend from $10^4$ to $10^6$ Pa s, although there are some indications that the upper limit could be extended somewhat. In any case, these are several orders of magnitude larger than the bulk viscosities encountered in usual hydrodynamics.

\noindent $\bullet$ In Sect. 6 we considered the {\it future} universe, extending from $t=0$ onwards. For definiteness we chose the value $\zeta_0= 10^5~$Pa s. We focused on the occurrence of a big rip singularity in the far future. The numerical values found in the earlier sections enabled us to make a quantitative estimate of the rip time $t_s$. With $\alpha$ defined as $\alpha=w+1$ we found that even the case $\zeta=\zeta_0=$const allows the big rip to occur, if $\alpha$ is negative, i.e., lying in the phantom region. This is the same kind of behavior as found earlier by Caldwell \cite{caldwell03} and others, in the nonviscous case. Of special interest is however the case $\zeta \propto \sqrt{\rho}$, where the fate of the universe is critically dependent on the magnitude of $\alpha$. If $\alpha<0$, the big rip is inevitable, similarly as above. If $\alpha >0$ (the quintessence region), the big rip can actually also occur if $\alpha$ is very small, less than about 0.005. This possibility of sliding through the phantom divide was actually pointed out several years ago \cite{Gorbunova}, but can now be better quantified. Typical rip times are found to lie roughly in the interval from 100 to 200 Gy.

%%%%%%%%%%%%%%%%%%%%%%%%%%%%%%%%%%%%%%%%%%
\vspace{6pt}

%%%%%%%%%%%%%%%%%%%%%%%%%%%%%%%%%%%%%%%%%%
%% optional
%\supplementary{The following are available online at www.mdpi.com/link, Figure S1: title, Table S1: title, Video S1: title.}

%%%%%%%%%%%%%%%%%%%%%%%%%%%%%%%%%%%%%%%%%%
\acknowledgments

\noindent{We thank Professor K\aa re Olaussen for valuable discussions.}

%%%%%%%%%%%%%%%%%%%%%%%%%%%%%%%%%%%%%%%%%%
%\authorcontributions{For research articles with several authors, a short paragraph specifying their individual contributions must be provided. The following statements should be used ``X.X. %and Y.Y. conceived and designed the experiments; X.X. performed the experiments; X.X. and Y.Y. analyzed the data; W.W. contributed reagents/materials/analysis tools; Y.Y. wrote the paper.'' %Authorship must be limited to those who have contributed substantially to the work reported.}

%%%%%%%%%%%%%%%%%%%%%%%%%%%%%%%%%%%%%%%%%%
 \conflictofinterests

\noindent{The authors declare no conflict of interest.}

%%%%%%%%%%%%%%%%%%%%%%%%%%%%%%%%%%%%%%%%%%
%% optional
%\abbreviations{The following abbreviations are used in this manuscript:\\

%\noindent MDPI: Multidisciplinary Digital Publishing Institute\\
%DOAJ: Directory of open access journals\\
%}

%%%%%%%%%%%%%%%%%%%%%%%%%%%%%%%%%%%%%%%%%%
\section*{Appendix}

\renewcommand{\theequation}{\mbox{\Alph{section}.\arabic{equation}}}
\appendix

\setcounter{equation}{0}
\section{Viscosity in expanding perfect fluids}
\label{App:A}
\noindent We first  point out a  simple though noteworthy property of an expanding universe consisting of many fluid components with homogeneous equations of state: the composite fluid when seen as {\it one single} fluid,  cannot itself have a homogeneous equation of state if $\zeta=0$. This statement is consistent also with that of Zimdahl \citep{zimdahl96}, who investigated this issue in more detail through a different approach.
Recall the energy-conservation equation for a inviscid fluid with equation of state $p=w\rho$ resulting from $\tensor{T}{^{\mu\nu}_{;\nu}}=0$:
\begin{equation}
\label{ApE_EnergyEq}
a\partial_a\rho+3(\rho+p)=0\phantom{000}\text{ with solution}\phantom{000}\rho(a)=\rho_0a^{-3(1+w)}.
\end{equation}
Now assume $\rho=\sum\rho_i$, where the components are distinguished by different homogeneous equations of state, for which the $w_i$s are all known. For simplicity we consider only two components (the argument holds also for more components),
\begin{equation}
\label{ApE_rho}
\rho=\rho_1+\rho_2,
\end{equation}
Inserting \eqref{ApE_rho} into the energy-conservation-equation in\eqref{ApE_EnergyEq} and summing, we find
\begin{equation}
\label{ApE_EnergyEq_2}
\sum_i [a\partial_a\rho_i+3(1+w_i)\rho_i]=0,
\end{equation}
from which we obtain after some simple manipulations
\begin{equation}
 a\partial_a\rho+3(1+w)\rho=3\sum_i(w-w_i)\rho_i.
\end{equation}
 Can we here choose an overall equation-of-state parameter $w$ such that the right-hand-side of the last equation  vanishes? If so, we would have constructed a phenomenological homogeneous (inviscid) fluid $\rho$ as the sum of two inviscid fluids possessing homogeneous equations of states. For this to happen, we must require
\begin{equation}
\label{ApE_EnergyEq_3}
w=\frac{\sum_i\rho_iw_i}{\rho}\phantom{00}\rightarrow\phantom{00}w\rho=\sum_i\rho_iw_i\rightarrow p=\sum_i p_i
\end{equation}
which  is nothing but Dalton's law for partial pressures,
If this requirement is not satisfied, there will be an additional contribution to the pressure balance which phenomenologically may be attributed to a viscosity. And it is clear that Eq.~(\ref{ApE_EnergyEq_3}) has to  be broken: taking $w_1$ and $w_2$  both constants but different from each other, it follows that  Eq.~\eqref{ApE_EnergyEq_3} cannot hold for $w=$const. The densities $\rho_1$ and $\rho_2$ evolve differently with respect to the scale factor $a$.  We thus see that even though $\rho_1$ and $\rho_2$ have homogeneous equations of state, $\rho=\rho_1+\rho_2$ cannot be seen as one effective fluid with a \textit{homogeneous equation of state}, without also introducing a phenomenological viscosity $\zeta$.

We can use the formalism above to give a more detailed derivation of the previous expression (\ref{EfectiveVisc}) for $\zeta$.  Equating the right-hand-side
  of Eq.~\eqref{ApE_EnergyEq_2} to $3\zeta\theta$ (cf. \eqref{C_EnergyEq}) we find
\begin{equation}
\label{E_RHS}
3\theta\zeta=3\sum_i(w-w_i)\rho_i
\end{equation}
which, by $\theta=\sqrt{24\pi G\rho}$, becomes
\begin{equation}
\label{E_visc}
\zeta=\frac{\sum_i(w-w_i)\rho_i}{\sqrt{24\pi G\rho}}.
\end{equation}
For a general multi-component fluid ($\rho=\rho_1+\rho_2+...+\rho_n$) of which two components $i=1,2$ are viewed as one component, the resulting expression for the viscosity will according to the above formula be
\begin{equation}
\label{E_visc}
\zeta=\frac{1}{\sqrt{24\pi G}}\left[w\sqrt{\rho}-w_1\frac{\rho_1}{\sqrt{\rho}}-w_2\frac{\rho_2}{\sqrt{\rho}}\right].
\end{equation}
Note that now $\rho=\rho_{comb}+\rho_3+...+\rho_n$, where $\rho_{comb}$ denotes the two components $i=1,2$ viewed as one component. The above expression may be rewritten as
\begin{equation}
\label{E_final}
\zeta=\frac{H_0}{8\pi G}\left[w-w_1\frac{\Omega_1(z)}{\Omega(z)}-w_2\frac{\Omega_2(z)}{\Omega(z)}\right]\sqrt{\Omega(z)},
\end{equation}
in agreement with Eq.~(\ref{EfectiveVisc}).

\setcounter{equation}{0}

\section{Comment on a universe filled solely with $\rho_\Lambda$}
\label{App:B}
Our final comment concerns the case where the only component in the  fluid $\rho$ is the cosmological constant ($\rho\rightarrow\rho_\Lambda$) obeying the equation
\begin{equation}
\label{C_EoS_Lambda}
p=-\rho_\Lambda .
\end{equation}
Then, since $\rho_\Lambda=\text{const}$, the energy-conservation-equation reduces to, if we reinstate the curvature parameter $k$,
\begin{equation}
\label{C_EnergyEq_Lambda}
\zeta_\Lambda\theta^2=\zeta_\Lambda\left(\frac{-k}{a^2}+\frac{\Lambda}{3}\right)=0,
\end{equation}
where the first Friedmann equation is used in the last equality. This leaves us with two options: (i)  $\Lambda=3k/a^2$, or (ii) $\zeta_\Lambda=0$. Imposing $k=0$ ,
one is left only with the last option. We can thus conclude that in flat space a cosmological fluid entirely consisting of a cosmological constant (i.e., $w=-1$), cannot be viscous.

%%%%%%%%%%%%%%%%%%%%%%%%%%%%%%%%%%%%%%%%%%
\bibliographystyle{mdpi}

%=====================================
% References, variant A: internal bibliography
%=====================================
\renewcommand\bibname{References}

%=====================================
% References, variant B: external bibliography
%=====================================
%\bibliography{your_external_BibTeX_file}

\end{document}